\documentclass[11pt]{article}
\usepackage{moriond}

\def\be{\begin{equation}}
\def\ee{\end{equation}}
\def\bea{\begin{eqnarray}}
\def\eea{\end{eqnarray}}

\newcommand{\Photo}{}
\newcommand{\m}{M_{H^{\pm}}}
\newcommand{\g}{\,\mbox{GeV}}
\newcommand{\la}{\lambda_1}
\newcommand{\lb}{\lambda_2}
\newcommand{\lc}{\lambda_3}

\newcommand{\lczp}{\lambda_{345}}
\newcommand{\rg}{R_{\gamma\gamma}}
\usepackage{amsmath}
\usepackage{amssymb}
\usepackage{amsthm}
\usepackage{amsfonts}

\begin{document}
\vspace*{4cm}
\title{2-PHOTON DECAY RATE OF THE SCALAR BOSON IN THE INERT DOUBLET MODEL}

\author{BOGUMI{\L}A \'{S}WIE\.{Z}EWSKA, MARIA KRAWCZYK}

\address{Faculty of Physics, University of Warsaw, Ho\.za 69, 00-681 Warsaw, Poland}

\maketitle\abstracts{Motivated by experimental hints on the possibility of deviating from the SM predictions for the 2-photon decay rate of the 125 GeV SM-like scalar boson $h$, an analysis of this rate in the framework of the Inert Doublet Model is presented. Regions in the parameter space where the 2-photon decay rate   is enhanced were found. The resulting regions are confronted with the allowed values of the Dark Matter (DM) particle's mass. Constraints on the masses of the charged scalar and the DM particle, and scalar couplings are presented.}

\section{Introduction}

The 2-photon decay of the Standard Model (SM)-like scalar boson, being one of the most important observational channels of the SM-like scalar  at the LHC, can also provide information about physics beyond the SM, because it is sensitive to the existence of new charged particles. Also existence of other new scalars can influence the signal strength because of the presence of the invisible decay channels. In this article we analyze the 2-photon decay rate in the framework of the Inert Doublet Model (IDM)~\cite{rg:2012}, see also~\cite{Posch:2010}, \cite{Arhrib:2012}, \cite{Stal:2013}, concentrating on the consequences of possible enhancement of this rate.

\section{Inert Doublet Model \label{idm}}
The IDM  is an extension of the Standard Model  with two scalar doublets $\Phi_S$ and $\Phi_D$, interacting according to the following potential~\cite{Krawczyk:2010}:
\be\label{pot}\renewcommand{\arraystretch}{1.2}
\begin{array}{rcl}
V&=&-\frac{1}{2}\left[m_{11}^{2}(\phi_{S}^{\dagger}\phi_{S})+m_{22}^{2}(\phi_{D}^{\dagger}\phi_{D})\right]+\frac{1}{2}\left[\lambda_{1}(\phi_{S}^{\dagger}\phi_{S})^{2}+\lambda_{2}(\phi_{D}^{\dagger}\phi_{D})^{2}\right]\\*
&&+\lambda_{3}(\phi_{S}^{\dagger}\phi_{S})(\phi_{D}^{\dagger}\phi_{D})+\lambda_{4}(\phi_{S}^{\dagger}\phi_{D})(\phi_{D}^{\dagger}\phi_{S})+\frac{1}{2}\lambda_{5}\left[(\phi_{S}^{\dagger}\phi_{D})^{2}+(\phi_{D}^{\dagger}\phi_{S})^{2}\right],
\end{array}
\ee
with all parameters real. This potential is symmetric under a discrete transformation $D$: $\Phi_S\to\Phi_S$, $\Phi_D\to-\Phi_D$. The Yukawa interactions  are also $D$-symmetric (only the $\Phi_S$ doublet couples to fermions) and so is the vacuum state (the so-called Inert vacuum): \renewcommand{\arraystretch}{.7}$\langle\phi_{S}\rangle=\left(\begin{array}{c}0\\v/\sqrt{2}\\\end{array}
\right)$, $\langle\phi_{D}\rangle=\left(\begin{array}{c}0\\0\\\end{array}\right)$, $v=246\g$. Thus, the symmetry $D$ is exact in the IDM.

The particle spectrum of the model consists of a SM-like scalar $h$ (originating from $\Phi_S$), which has tree-level couplings to fermions and gauge bosons just like the SM scalar boson, and four dark (inert) scalars: $H$, $A$ and $H^{\pm}$ (coming from $\Phi_D$), which do not couple to fermions. Due to the $D$-parity conservation the lightest $D$-odd (dark) particle is stable and thus constitutes a Dark Matter (DM) candidate, given that is electrically neutral. In this work we assume that $H$ is the DM candidate. It has been shown that the DM coming from the IDM can be consistent with the WMAP observations only in three mass regimes:  $M_{\textrm{DM}}\lesssim 10\g$, $40\lesssim M_{\textrm{DM}}\lesssim 80\g$ or $M_{\textrm{DM}}\gtrsim 500\g$~\cite{Dolle:2009}. We will confront this regions with the results of the $\rg$ analysis.

\subsection{Constraints\label{warunki}}

The numerical results presented in Section~\ref{wyniki} are obtained by scanning randomly the parameter space of the IDM taking into account the following constraints:
\begin{description}
\item[Vacuum stability:] For a stable vacuum state to exist it is necessary that: $\la>0,\ \lb>0,\ \lc+\sqrt{\la\lb}>0,\ \lczp+\sqrt{\la\lb}>0.$
\item[Perturbative unitarity:] We demand that the eigenvalues $\Lambda_i$ of the high energy scattering matrix of the scalar sector fulfill the condition: $|\Lambda_i|<8\pi$.
\item[Existence of the Inert vacuum:] For the Inert state to be the global minimum of the potential it is necessary that it is a minimum, i.e., the scalars' masses squared are positive and that its energy is lower than the energy of coexisting minima. The latter is assured by: $\frac{m_{11}^2}{\sqrt{\la}}>\frac{m_{22}^2}{\sqrt{\lb}}$, which when combined with the unitarity constraints on $\lb$ and the measured mass of the scalar boson $M_h=125\g$ gives the condition~\cite{Swiezewska:2012}
\be 
m_{22}^2\lesssim 9\cdot 10^4 \g^2.\label{inert-vac}
\ee
\item[Electroweak Precision Tests (EWPT):] We require that the values of $S$ and $T$ parameters lie within $2\sigma$ ellipses around the central values: $S=0.03\pm0.09$, $T=0.07\pm0.08$ (with correlation equal to 87\%)~\cite{Nakamura:2010}.
\item LEP bounds: We impose the LEP bounds~\cite{Gustafsson:2009} on scalar masses: $\m +M_H>M_{W},\  \m+ M_A>M_W,\ M_H+M_A >M_Z,\ 2\m > M_Z,\ \m>70\g$
and exclude the region where simultaneously: $M_H< 80\g,\  M_A< 100\g\ \textrm{and}\  M_A - M_H> 8\g$.
\item[LHC data:] We set  $M_h=125\g$~\cite{cms:2012,atlas:2012}.
\item[$H$ as DM candidate:] As $H$ is supposed to be the DM candidate, we assume $M_H<M_A,\ \m$.
\end{description}

\section{2-photon decay rate of the scalar boson}
The 2-photon decay rate of the SM-like scalar boson is defined as follows:
\be
R_{\gamma \gamma}:=\frac{\sigma(pp\to h\to \gamma\gamma)^{\textrm{IDM}}}{\sigma(pp\to h\to \gamma\gamma)^{\textrm  {SM}}}
\approx\frac{\Gamma(h\to\gamma\gamma)^{\textrm {IDM}}}{\Gamma(h\to\gamma\gamma)^{\textrm {SM}}}\frac{\Gamma(h)^{\textrm {SM}}}{\Gamma(h)^{\textrm {IDM}}},
\ee
where the facts that the main production channel is via gluon fusion and that $h$ couples (at the tree-level) to SM particles  like SM scalar boson, so $\sigma(gg\to h)^{\mathrm{IDM}}=\sigma(gg\to h)^{\mathrm{SM}}$, have been taken into account. It can be seen that deviations from $\rg=1$ can be caused by:
\begin{description}
\item[Modifications of the total decay width of $h$.] 
The total decay width $\Gamma(h)^{\textrm {IDM}}$ is modified with respect to the SM value dominantly due to the existence of the invisible decay channels: $h\to HH$ or $h\to AA$. (Loop induced modifications are negligible in comparison with the SM total width of $h$.) 
\item[Modifications of the partial decay width to two photons.] The source of modifications of the partial decay width given by the formula:
\be
\Gamma(h\to\gamma\gamma)^{\textrm{IDM}}=\frac{G_F\alpha^2M_h^3}{128\sqrt{2}\pi^3}\left | \mathcal{A}^{\textrm{SM}}+\frac{2\m^2+m_{22}^2}{2\m^2}A_0\left(\frac{4\m^2}{M_h^2}\right)\right |^2\label{rg}
\ee
is the charged scalar loop contribution, which  can interfere either constructively or destructively with the SM term $\mathcal{A}^{\mathrm{SM}}$ (for the definition of $\mathcal{A}^{\textrm{SM}}$ and $A_0$ see~\cite{Djouadi:2005}). 
\end{description}
The final value of $\rg$ is a result of interplay between these two factors.

\section{Results\label{wyniki}}
\subsection{Invisible channels open\label{results-inv-open}}
\begin{figure}[ht]
\centering
\includegraphics[width=0.4\textwidth]{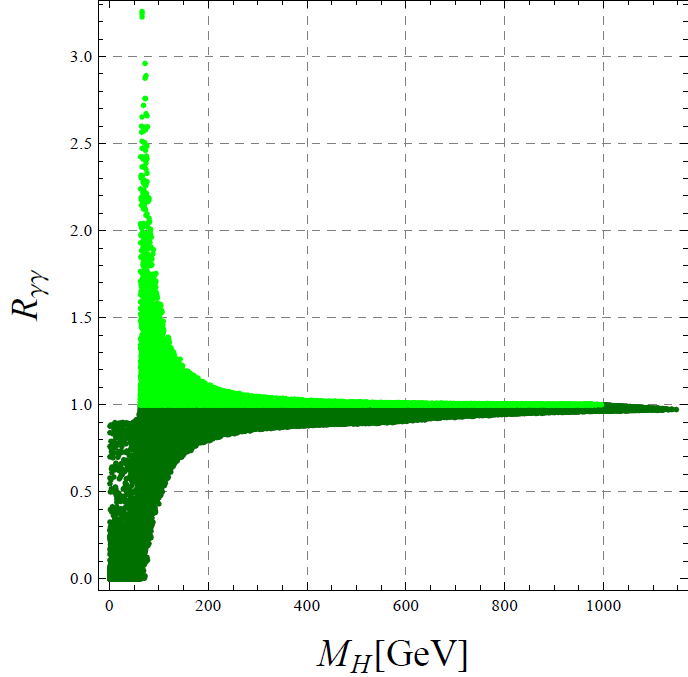}\hspace{1cm}
\includegraphics[width=0.4\textwidth]{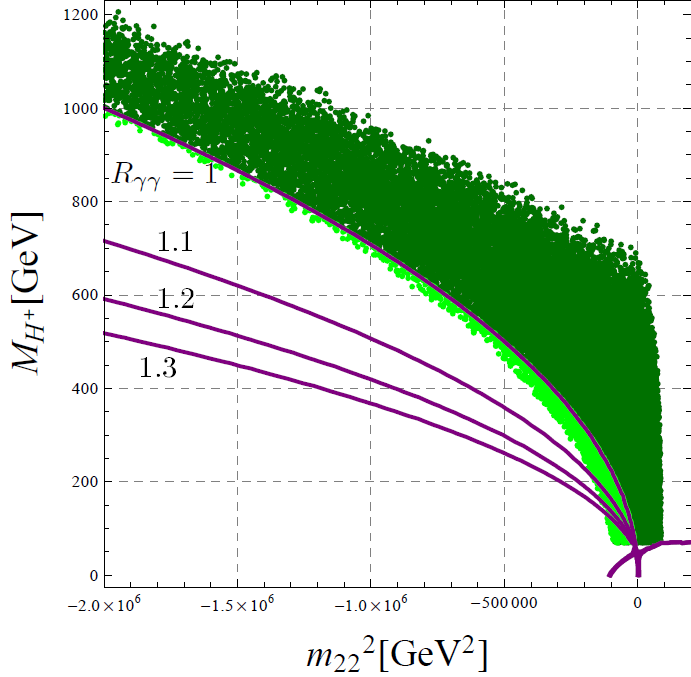}
\caption{Left panel: $\rg$ as a function of $M_H$ for $-2\cdot10^6 \g^2 \leqslant m_{22}^2\leqslant 9\cdot 10^4 \g^2$. Right panel: Region allowed in the $(m_{22}^2,\m)$ plane by the constraints from Section~\ref{warunki} together with curves representing constant values of $\rg$. Points with $\rg>1$ ($\rg<1$) are displayed in light (dark) green/gray. \label{fig}
}
\end{figure}

In Fig.~\ref{fig} (left panel) we present the dependence of $\rg$ on $M_H$. It can be seen that when the invisible channel $h\to HH$ is open (for $M_H<M_h/2\approx62.5\g$), enhancing $\rg$ is impossible~\cite{Arhrib:2012}. Nonetheless, for this case any value of $\rg$ up to around $0.9$ can be obtained.
Note, that  since $M_H<M_A,\m$, enhanced 2-photon decay rate of $h$ can be observed only if $M_H, M_A, \m>M_h/2$. Therefore, if $\rg>1$ is confirmed, light DM is excluded.

\subsection{Invisible channels closed}
When the invisible decay channels are kinematically closed Eq.~(\ref{rg}) reduces to:
\be
R_{\gamma \gamma}=\frac{\Gamma(h\to\gamma\gamma)^{\textrm {IDM}}}{\Gamma(h\to\gamma\gamma)^{\textrm {SM}}}.
\ee
In this case we solved the inequality $\rg>1$   analytically. As a result we found that the 2-photon decay rate can be enhanced only for:
\begin{align}
m_{22}^2<&-2\m^2<-9.8\cdot10^3\g^2\ \quad \textrm{or}\label{constr}\\
m_{22}^2>&\frac{a M_h^2}{1-\left(\frac{2 \m}{M_h}\right)^2\arcsin^2\left(\frac{M_h}{2 \m}\right)}-2\m^2\gtrsim1.8\cdot 10^5\g^2\label{destr},
\end{align}
where $a=\textrm{Re}\mathcal{A}^{\textrm{SM}}$ and for obtaining the numerical values the LEP bound on the charged scalar mass $\m>70\g$ has been used. These two options correspond to constructive (Eq.~(\ref{constr})) or destructive (Eq.~(\ref{destr})) interference between the charged scalar loop and the SM contribution. The latter case is excluded by the condition for the existence of the Inert vacuum, Eq.~(\ref{inert-vac}). The surviving condition $m_{22}^2<-2\m^2$ can be translated to the condition for the coupling between $h$ and $H^+H^-$ as $\lc<0$, see also~\cite{Arhrib:2012}.

In Fig.~\ref{fig} (right panel) we present the region allowed by the constraints described in Section~\ref{warunki} in the $(m_{22}^2,\m)$ plane. The points for which $\rg>1$ are displayed in light green/gray. In addition, curves representing fixed values of $\rg$ (calculated for the invisible channels closed) are shown. It can be observed, that the 2-photon decay rate can be enhanced for any value of the charged scalar mass. However, if  $\rg$ is demanded to be sharply greater than 1, then an upper limit on $\m$ arises. For example if $\rg>1.3$, then $\m, M_H<135\g$. Joining these results with the results from Section~\ref{results-inv-open} and the LEP bound on $\m$ it can be seen that  $\rg>1.3$ implies:
\be
70\g<\m<135\g, \quad 62.5\g<M_H<135\g.
\ee
 The bound mentioned above  leaves only the medium mass regime of the DM viable.
 
 In similar way the couplings between $h$ and $H^{\pm}$ or $H$ can be constrained. Demanding $\rg>1$, we get the condition $\lc,\lczp<0$, while $\rg>1.3$ implies:
 \be
 -1.46<\lambda_3,\lambda_{345} <-0.24.
 \ee
\section{Summary}
We have analyzed the 2-photon decay rate of the SM-like scalar boson $h$ within the framework of the IDM. Taking into account  a number of constraints 
we have found the regions in the parameter space where this rate can be enhanced with respect to the SM case. We conclude that it is possible only if $M_H,M_A,\m<M_h/2$. Furthermore, if a lower bound, which is greater than one, is set on $\rg$, an upper bound on the masses of the charged scalar and the DM particle, and their couplings to $h$ arises. For example for the case with $\rg>1.3$ the following bounds are obtained: $70\g<\m<135\g,\ 62.5\g<M_H<135\g$, $-1.46<\lambda_3,\lambda_{345} <-0.24$.

\section*{Acknowledgments}
B\'{S} would like to thank the \textit{Recontres de Moriond EW 2013} organizers for the opportunity to present this work and for their financial support. This work was supported in part by a grant NCN OPUS 2012/05/B/ST2/03306 (2012-
2016).

\section*{References}
\bibliographystyle{unsrt} 
\bibliography{biblio}

\begin{thebibliography}{10}

\bibitem{rg:2012}
Bogumila Swiezewska and Maria Krawczyk.
\newblock {Diphoton rate in the Inert Doublet Model with a 125 GeV Higgs
  boson}.
\newblock 2012.

\bibitem{Posch:2010}
Paul Posch.
\newblock {Enhancement of $h \to \gamma \gamma$ in the Two Higgs Doublet Model
  Type I}.
\newblock {\em Phys.Lett.}, B696:447--453, 2011.

\bibitem{Arhrib:2012}
Abdesslam Arhrib, Rachid Benbrik, and Naveen Gaur.
\newblock {$H\to \gamma \gamma$ in Inert Higgs Doublet Model}.
\newblock {\em Phys.Rev.}, D85:095021, 2012.

\bibitem{Stal:2013}
A.~Goudelis, B.~Herrmann, and O.~St{\aa}l.
\newblock {Dark matter in the Inert Doublet Model after the discovery of a
  Higgs-like boson at the LHC}.
\newblock 2013.

\bibitem{Krawczyk:2010}
I.F. Ginzburg, K.A. Kanishev, M.~Krawczyk, and D.~Soko³owska.
\newblock {Evolution of Universe to the present inert phase}.
\newblock {\em Phys.Rev.}, D82:123533, 2010.

\bibitem{Dolle:2009}
Ethan~M. Dolle and Shufang Su.
\newblock {The Inert Dark Matter}.
\newblock {\em Phys.Rev.}, D80:055012, 2009.

\bibitem{Swiezewska:2012}
Bogumila Swiezewska.
\newblock {Yukawa independent constraints for 2HDMs with a 125 GeV Higgs
  boson}.
\newblock 2012.

\bibitem{Nakamura:2010}
K.~Nakamura et~al.
\newblock {Review of particle physics}.
\newblock {\em J.Phys.G}, G37:075021, 2010.

\bibitem{Gustafsson:2009}
Erik Lundstrom, Michael Gustafsson, and Joakim Edsjo.
\newblock {The Inert Doublet Model and LEP II Limits}.
\newblock {\em Phys.Rev.}, D79:035013, 2009.

\bibitem{cms:2012}
Serguei Chatrchyan et~al.
\newblock {Observation of a new boson at a mass of 125 GeV with the CMS
  experiment at the LHC}.
\newblock {\em Phys.Lett.}, B716:30--61, 2012.

\bibitem{atlas:2012}
Georges Aad et~al.
\newblock {Observation of a new particle in the search for the Standard Model
  Higgs boson with the ATLAS detector at the LHC}.
\newblock {\em Phys.Lett.}, B716:1--29, 2012.

\bibitem{Djouadi:2005}
Abdelhak Djouadi.
\newblock {The Anatomy of electro-weak symmetry breaking. II. The Higgs bosons
  in the minimal supersymmetric model}.
\newblock {\em Phys.Rept.}, 459:1--241, 2008.

\end{thebibliography}

\end{document}